\documentclass[aps,prl,showpacs,twocolumn,groupedaddress,superscriptaddress]{revtex4}  
\usepackage{graphicx}
\usepackage{dcolumn}
\usepackage{bm}
\usepackage{epsfig}
\usepackage{multirow}
\begin{document}
%
%
\newcommand{\comm}[1]{\mbox{\mbox{\textup{#1}}}}
\newcommand{\subs}[1]{\mbox{\scriptstyle \mathit{#1}}}
\newcommand{\subss}[1]{\mbox{\scriptscriptstyle \mathit{#1}}}
\newcommand{\Frac}[2]{\mbox{\frac{\displaystyle{#1}}{\displaystyle{#2}}}}
\newcommand{\LS}[1]{\mbox{_{\scriptstyle \mathit{#1}}}}
\newcommand{\US}[1]{\mbox{^{\scriptstyle \mathit{#1}}}}
\def\gsim{\mathrel{\rlap{\raise.4ex\hbox{$>$}} {\lower.6ex\hbox{$\sim$}}}}
\def\lsim{\mathrel{\rlap{\raise.4ex\hbox{$<$}} {\lower.6ex\hbox{$\sim$}}}}
\renewcommand{\arraystretch}{1.3}
\newcommand{\Edep}{\mbox{E_{\mathit{dep}}}}
\newcommand{\Ebeam}{\mbox{E_{\mathit{beam}}}}
\newcommand{\Exrc}{\mbox{E_{\mathit{rec}}^{\mathit{exp}}}}
\newcommand{\Emrc}{\mbox{E_{\mathit{rec}}^{\mathit{sim}}}}
\newcommand{\Evis}{\mbox{E_{\mathit{vis}}}}
\newcommand{\Edepi}{\mbox{E_{\mathit{dep,i}}}}
\newcommand{\Evisi}{\mbox{E_{\mathit{vis,i}}}}
\newcommand{\Exrci}{\mbox{E_{\mathit{rec,i}}^{\mathit{exp}}}}
\newcommand{\Emrci}{\mbox{E_{\mathit{rec,i}}^{\mathit{sim}}}}
\newcommand{\Etmis}{\mbox{E_{\mathit{t,miss}}}}
%
%
\newcommand{\lt}{\mbox{$<$}}
\newcommand{\gt}{\mbox{$>$}}
\newcommand{\lte}{\mbox{$\le$}}
\newcommand{\gte}{\mbox{$\ge$}}

\newcommand{\xa}{\mbox{$x_{a}$}}
\newcommand{\xb}{\mbox{$x_{b}$}}
\newcommand{\xp}{\mbox{$x_{p}$}}
\newcommand{\xpb}{\mbox{$x_{\bar{p}}$}}

\newcommand{\alphas}{\mbox{$\alpha_{s}$}}
\newcommand{\partt}{\mbox{$\partial^{t}$}}
\newcommand{\partd}{\mbox{$\partial_{t}$}}
\newcommand{\parmut}{\mbox{$\partial^{\mu}$}}
\newcommand{\parmud}{\mbox{$\partial_{\mu}$}}
\newcommand{\parnut}{\mbox{$\partial^{\nu}$}}
\newcommand{\parnud}{\mbox{$\partial_{\nu}$}}
\newcommand{\Amut}{\mbox{$A^{\mu}$}}
\newcommand{\Amud}{\mbox{$A_{\mu}$}}
\newcommand{\Anut}{\mbox{$A^{\nu}$}}
\newcommand{\Anud}{\mbox{$A_{\nu}$}}
\newcommand{\Fmunut}{\mbox{$F^{\mu\nu}$}}
\newcommand{\Fmunud}{\mbox{$F_{\mu\nu}$}}
\newcommand{\Gmut}{\mbox{$\gamma^{\mu}$}}
\newcommand{\Gmud}{\mbox{$\gamma_{\mu}$}}
\newcommand{\Gnut}{\mbox{$\gamma^{\nu}$}}
\newcommand{\Gnud}{\mbox{$\gamma_{\nu}$}}
\newcommand{\pvecsq}{\mbox{$\vec{p}^{\:2}$}}
\newcommand{\bigsum}{\mbox{$\displaystyle{\sum}$}}
%
\newcommand{\Lzero}{Level \O}
\newcommand{\Lone}{Level $1$}
\newcommand{\Ltwo}{Level $2$}
\newcommand{\Lhalf}{Level $1.5$}
%
\newcommand{\bzero}{\mbox{\comm{B\O}}}
\newcommand{\dzero}{\mbox{D\O}}
\newcommand{\dzerosm}{\mbox{\comm{$\scriptsize{D\O}$}}}
\newcommand{\runb}{Run~$1$B}
\newcommand{\runa}{Run~$1$A}
\newcommand{\runone}{Run~$1$}
\newcommand{\runtwo}{Run~$2$}
\newcommand{\dzpjet}{\textsc{D{\O}Pjet}}
\newcommand{\y}{\mbox{$y$}}
\newcommand{\z}{\mbox{$z$}}
\newcommand{\px}{\mbox{$p_{x}$}}
\newcommand{\py}{\mbox{$p_{y}$}}
\newcommand{\pz}{\mbox{$p_{z}$}}
\newcommand{\ex}{\mbox{$E_{x}$}}
\newcommand{\ey}{\mbox{$E_{y}$}}
\newcommand{\ez}{\mbox{$E_{z}$}}
\newcommand{\et}{\mbox{$E_{T}$}}
\newcommand{\etprime}{\mbox{$E_{T}^{\prime}$}}
\newcommand{\etone}{\mbox{$E_{T}^{\mathrm{1}}$}}
\newcommand{\ettwo}{\mbox{$E_{T}^{\mathrm{2}}$}}
\newcommand{\etlj}{\mbox{$E_{T}^{\subs{lj}}$}}
\newcommand{\etmax}{\mbox{$E_{T}^{max}$}}
\newcommand{\etcand}{\mbox{$E_{T}^{\subs{cand}}$}}
\newcommand{\etup}{\mbox{$E_{T}^{\subs{up}}$}}
\newcommand{\etdown}{\mbox{$E_{T}^{\subs{down}}$}}
\newcommand{\jet}{\mbox{$E_{T}^{\subs{jet}}$}}
\newcommand{\cet}{\mbox{$E_{T}^{\subs{cell}}$}}
\newcommand{\jetvec}{\mbox{$\vec{E}_{T}^{\subs{jet}}$}}
\newcommand{\cetvec}{\mbox{$\vec{E}_{T}^{\subs{cell}}$}}
\newcommand{\jevec}{\mbox{$\vec{E}^{\subs{jet}}$}}
\newcommand{\cevec}{\mbox{$\vec{E}^{\subs{cell}}$}}
\newcommand{\etfr}{\mbox{$f_{E_{T}}$}}
\newcommand{\aveet}{\mbox{$\langle\et\rangle$}}
\newcommand{\nj}{\mbox{$n_{j}$}}
\newcommand{\ptrel}{\mbox{$p_{T}^{rel}$}}
\newcommand{\etad}{\mbox{$\eta_{d}$}}           
\newcommand{\peta}{\mbox{$\eta$}}               
\newcommand{\aeta}{\mbox{$|\eta|$}}             
\newcommand{\ifb}{fb$^{-1}$}
\newcommand{\ipb}{pb$^{-1}$}
\newcommand{\inb}{nb$^{-1}$}
\newcommand{\met}{\mbox{${\hbox{$E$\kern-0.63em\lower-.18ex\hbox{/}}}_{T}$}}
\newcommand{\metvec}{\mbox{${\hbox{$\vec{E}$\kern-0.63em\lower-.18ex\hbox{/}}}_{T}\,$}}
\newcommand{\metx}{\mbox{${\hbox{$E$\kern-0.63em\lower-.18ex\hbox{/}}}_{x}\,$}}
\newcommand{\mety}{\mbox{${\hbox{$E$\kern-0.63em\lower-.18ex\hbox{/}}}_{y}\,$}}
\newcommand{\het}{\mbox{$\vec{\mathcal{H}}_{T}$}}
\newcommand{\hetsc}{\mbox{$\mathcal{H}_{T}$}}
\newcommand{\zvrt}{\mbox{$Z$}}
\newcommand{\zcut}{\mbox{$|\zvrt| < 50$}}
\newcommand{\mtwo}{\mbox{$\mathcal{M}_{2}$}}
\newcommand{\mthree}{\mbox{$\mathcal{M}_{3}$}}
\newcommand{\mfour}{\mbox{$\mathcal{M}_{4}$}}
\newcommand{\msix}{\mbox{$\mathcal{M}_{6}$}}
\newcommand{\mn}{\mbox{$\mathcal{M}_{n}$}}
\newcommand{\R}{\mbox{$R_{\subss{MTE}}$}}
\newcommand{\invR}{\mbox{$1/R_{\subss{MTE}}$}}
\newcommand{\eemf}{\mbox{$\varepsilon_{\subss{EMF}}$}}
\newcommand{\echf}{\mbox{$\varepsilon_{\subss{CHF}}$}}
\newcommand{\ehcf}{\mbox{$\varepsilon_{\subss{HCF}}$}}
\newcommand{\eglob}{$\mbox{\varepsilon_{\subs{glob}}$}}
\newcommand{\emte}{\mbox{$\varepsilon_{\subss{MTE}}$}}
\newcommand{\ezvrt}{\mbox{$\varepsilon_{\subss{Z}}$}}
\newcommand{\etot}{\mbox{$\varepsilon_{\subs{tot}}$}}
\newcommand{\Bprime}{\mbox{$\comm{B}^{\prime}$}}
\newcommand{\Nsurv}{\mbox{$N_{\subs{surv}}$}}
\newcommand{\Nfail}{\mbox{$N_{\subs{fail}}$}}
\newcommand{\Ntot}{\mbox{$N_{\subs{tot}}$}}
\newcommand{\p}[1]{\mbox{$p_{#1}$}}
\newcommand{\ep}[1]{\mbox{$\Delta p_{#1}$}}
\newcommand{\delr}{\mbox{$\Delta R$}}
\newcommand{\deleta}{\mbox{$\Delta\eta$}}
\newcommand{\cafone}{{\sc Cafix 5.1}}
\newcommand{\caftwo}{{\sc Cafix 5.2}}
\newcommand{\delphi}{\mbox{$\Delta\varphi$}}
\newcommand{\rphi}{\mbox{$r-\varphi$}}
\newcommand{\etaphi}{\mbox{$\eta-\varphi$}}
\newcommand{\etatphi}{\mbox{$\eta\times\varphi$}}
\newcommand{\Rjet}{\mbox{$R_{jet}$}}
\newcommand{\jphi}{\mbox{$\varphi_{\subs{jet}}$}}
\newcommand{\gphi}{\mbox{$\varphi_{\subs{\gamma}}$}}
\newcommand{\ceta}{\mbox{$\eta^{\subs{cell}}$}}
\newcommand{\cphi}{\mbox{$\phi^{\subs{cell}}$}}
\newcommand{\inlum}{\mbox{$\mathcal{L}$}}
\newcommand{\gm}{\mbox{$\gamma$}}
\newcommand{\Rjj}{\mbox{$\mathbf{R}_{\subs{jj}}$}}
\newcommand{\Rgj}{\mbox{$\mathbf{R}_{\subs{\gamma j}}$}}
\newcommand{\Rmathcal}{\mbox{$\mathcal{R}$}}
\newcommand{\etv}{\mbox{$\vec{E}_{T}$}}
\newcommand{\nvec}{\mbox{$\hat{\vec{n}}$}}
\newcommand{\eprime}{\mbox{$E^{\prime}$}}
\newcommand{\aveprime}{\mbox{$\bar{E}^{\prime}$}}
\newcommand{\geta}{\mbox{$\eta_{\gm}$}}
\newcommand{\jeta}{\mbox{$\eta_{\subs{jet}}$}}
\newcommand{\cjeta}{\mbox{$\eta_{\subs{jet}}^{\subss{CEN}}$}}
\newcommand{\fjeta}{\mbox{$\eta_{\subs{jet}}^{\subss{FOR}}$}}
\newcommand{\cjphi}{\mbox{$\varphi_{\subs{jet}}^{\subss{CEN}}$}}
\newcommand{\fjphi}{\mbox{$\varphi_{\subs{jet}}^{\subss{FOR}}$}}
\newcommand{\etcut}{\mbox{$E_{T}^{\subs{cut}}$}}
\newcommand{\etg}{\mbox{$E_{T}^{\gm}$}}
\newcommand{\cenet}{\mbox{$E_{T}^{\subss{CEN}}$}}
\newcommand{\foret}{\mbox{$E_{T}^{\subss{FOR}}$}}
\newcommand{\cenen}{\mbox{$E^{\subss{CEN}}$}}
\newcommand{\foren}{\mbox{$E^{\subss{FOR}}$}}
\newcommand{\ejtptc}{\mbox{$E^{\subs{ptcl}}_{\subs{jet}}$}}
\newcommand{\ejtmes}{\mbox{$E^{\subs{meas}}_{\subs{jet}}$}}
\newcommand{\AIDA}{{\sc AIDA}}
\newcommand{\RECO}{{\sc Reco}}
\newcommand{\PYTHIA}{{\sc Pythia}}
\newcommand{\HERWIG}{{\sc Herwig}}
\newcommand{\JETRAD}{{\sc Jetrad}}
\newcommand{\CTone}{\mbox{$|\eta|<0.4}$}
\newcommand{\CTtwo}{\mbox{$0.4\leq|\eta|<0.8$}}
\newcommand{\ICone}{\mbox{$0.8\leq|\eta|<1.2$}}
\newcommand{\ICtwo}{\mbox{$1.2\leq|\eta|<1.6$}}
\newcommand{\FWone}{\mbox{$1.6\leq|\eta|<2.0$}}
\newcommand{\FWtwo}{\mbox{$2.0\leq|\eta|<2.5$}}
\newcommand{\FWthr}{\mbox{$2.5\leq|\eta|<3.0$}}
\newcommand{\LCTone}{\mbox{$|\eta|<0.5$}}
\newcommand{\LCTtwo}{\mbox{$0.5\leq|\eta|<1.0$}}
\newcommand{\LICone}{\mbox{$1.0\leq|\eta|<1.5$}}
\newcommand{\LICtwo}{\mbox{$1.5\leq|\eta|<2.0$}}
\newcommand{\LFWone}{\mbox{$2.0\leq|\eta|<3.0$}}
\newcommand{\CSone}{\mbox{$|\eta|<0.5$}}
\newcommand{\CStwo}{\mbox{$0.5\leq|\eta|<1.0$}}
\newcommand{\CSthr}{\mbox{$1.0\leq|\eta|<1.5$}}
\newcommand{\CSfou}{\mbox{$1.5\leq|\eta|<2.0$}}
\newcommand{\CSfiv}{\mbox{$2.0\leq|\eta|<3.0$}}
\newcommand{\sigA}{\mbox{$\sigma_{\subss{\!A}}$}}
\newcommand{\sigASS}{\mbox{$\sigma_{\subss{A}}^{\subss{SS}}$}}
\newcommand{\sigAOS}{\mbox{$\sigma_{\subss{A}}^{\subss{OS}}$}}
\newcommand{\sigZ}{\mbox{$\sigma_{\subss{Z}}$}}
\newcommand{\sige}{\mbox{$\sigma_{\subss{E}}$}}
\newcommand{\siget}{\mbox{$\sigma_{\subss{\et}}$}}
\newcommand{\sigetone}{\mbox{$\sigma_{\subs{\etone}}$}}
\newcommand{\sigettwo}{\mbox{$\sigma_{\subs{\ettwo}}$}}
\newcommand{\rcal}{\mbox{$R_{\subs{cal}}$}}
\newcommand{\zcal}{\mbox{$Z_{\subs{cal}}$}}
\newcommand{\Runf}{\mbox{$R_{\subs{unf}}$}}
\newcommand{\Rsep}{\mbox{$\mathcal{R}_{sep}$}}
\newcommand{\etal}{{\it et al.}}
\newcommand{\ppbar}{\mbox{$p\overline{p}$}}
\newcommand{\pp}{\mbox{$pp$}}
\newcommand{\qqbar}{\mbox{$q\overline{q}$}}
\newcommand{\ccbar}{\mbox{$c\overline{c}$}}
\newcommand{\bbbar}{\mbox{$b\overline{b}$}}
\newcommand{\ttbar}{\mbox{$t\overline{t}$}}

\newcommand{\bbj}{\mbox{$b\overline{b}j$}}
\newcommand{\bbjj}{\mbox{$b\overline{b}jj$}}
\newcommand{\ccjj}{\mbox{$c\overline{c}jj$}}

\newcommand{\bb}{\mbox{$b\overline{b}j(j)$}}
\newcommand{\cc}{\mbox{$c\overline{c}j(j)$}}

\newcommand{\hboson}{\mbox{$\mathit{h}$}}
\newcommand{\Hboson}{\mbox{$\mathit{H}$}}
\newcommand{\Aboson}{\mbox{$\mathit{A}$}}
\newcommand{\zboson}{\mbox{$\mathit{Z}$}}
\newcommand{\zb}{\mbox{$\mathit{Zb}$}}
\newcommand{\bh}{\mbox{$\mathit{bh}$}}
\newcommand{\btag}{\mbox{$\mathit{b}$}}

\newcommand{\hsm}{\mbox{$h_{SM}$}}
\newcommand{\hmssm}{\mbox{$h_{MSSM}$}}

\newcommand{\prot}{\mbox{$p$}}
\newcommand{\pbar}{\mbox{$\overline{p}$}}
\newcommand{\pt}{\mbox{$p_{T}$}}
\newcommand{\xnot}{\mbox{$X_{0}$}}
\newcommand{\Znot}{\mbox{$Z^{0}$}}
\newcommand{\Wpm}{\mbox{$W^{\pm}$}}
\newcommand{\Wplus}{\mbox{$W^{+}$}}
\newcommand{\Wminus}{\mbox{$W^{-}$}}
\newcommand{\lamb}{\mbox{$\lambda$}}
\newcommand{\nhatbf}{\mbox{$\hat{\mathbf{n}}$}}
\newcommand{\pbf}{\mbox{$\mathbf{p}$}}
\newcommand{\xbf}{\mbox{$\mathbf{x}$}}
\newcommand{\jbf}{\mbox{$\mathbf{j}$}}
\newcommand{\Ebf}{\mbox{$\mathbf{E}$}}
\newcommand{\Bbf}{\mbox{$\mathbf{B}$}}
\newcommand{\Abf}{\mbox{$\mathbf{A}$}}
\newcommand{\Rbf}{\mbox{$\mathbf{R}$}}
\newcommand{\nablabf}{\mbox{$\mathbf{\nabla}$}}
\newcommand{\rarrow}{\mbox{$\rightarrow$}}
\newcommand{\slashp}{\mbox{$\not \! p \,$}}
\newcommand{\slashk}{\mbox{$\not \! k$}}
\newcommand{\slasha}{\mbox{$\not \! a$}}
\newcommand{\slashA}{\mbox{$\! \not \! \! A$}}
\newcommand{\slashpar}{\mbox{$\! \not \! \partial$}}
\newcommand{\intdouble}{\mbox{$\int\!\!\int$}}
\newcommand{\MRSTGU}{MRSTg$\uparrow$}
\newcommand{\MRSTGD}{MRSTg$\downarrow$}
%
\newcommand{\Due}{\mbox{$D_{\mathrm{ue}}$}}
\newcommand{\Dth}{\mbox{$D_{\Theta}$}}
\newcommand{\Dof}{\mbox{$D_{\mathrm{O}}$}}
\newcommand{\zbl}{\texttt{ZERO BIAS}}
\newcommand{\mbl}{\texttt{MIN BIAS}}
\newcommand{\mbll}{\texttt{MINIMUM BIAS}}
\newcommand{\nue}{\mbox{$\nu_{e}$}}
\newcommand{\num}{\mbox{$\nu_{\mu}$}}
\newcommand{\nut}{\mbox{$\nu_{\tau}$}}
\newcommand{\mycs}{\mbox{$d^{\,2}\sigma/(d\et d\eta)$}}
\newcommand{\mycsav}{\mbox{$\langle \mycs \rangle$}}
\newcommand{\tdcs}{\mbox{$d^{\,3}\sigma/d\et d\eta_{1} d\eta_{2}$}}
\newcommand{\tdcsav}{\mbox{$\langle d^{\,3}\sigma/d\et d\eta_{1} d\eta_{2} \rangle$}}
\newcommand{\tanb}{$\tan\beta$}
\newcommand{\cotb}{$\cot\beta$}

\newcommand{\rstev}{\mbox{$\rs = \T{1.8}$}}
\newcommand{\rssps}{\mbox{$\rs = \T{0.63}$}}
\newcommand{\XX}{\mbox{$\, \times \,$}}
\newcommand{\AP}{\mbox{${\rm \bar{p}}$}}
\newcommand{\SU}{\mbox{$<\! |S|^2 \!>$}}
\newcommand{\ET}{\mbox{$E_{T}$}}
\newcommand{\HT}{\mbox{$S_{{\rm {\sl T}}}$} }
\newcommand{\PT}{\mbox{$p_{t}$}}
\newcommand{\DP}{\mbox{$\Delta\phi$}}
\newcommand{\DR}{\mbox{$\Delta R$}}
\newcommand{\DE}{\mbox{$\Delta\eta$}}
\newcommand{\DEP}{\mbox{$\Delta\eta_{c}$}}
\newcommand{\PH}{\mbox{$\phi$}}
\newcommand{\EA}{\mbox{$\eta$} }
\newcommand{\EAJ}{\mbox{\EA(jet)}}
\newcommand{\AEA}{\mbox{$|\eta|$}}
\newcommand{\Ge}[1]{\mbox{#1 GeV}}
\newcommand{\T}[1]{\mbox{#1 TeV}}
\newcommand{\x}{\cdot}
\newcommand{\ra}{\rightarrow}
\def\D0{D\O}
\def\ETmiss{{\rm {\mbox{$E\kern-0.57em\raise0.19ex\hbox{/}_{T}$}}}}
\newcommand{\mb}{\mbox{mb}}
\newcommand{\nb}{\mbox{nb}}
\newcommand{\rs}{\mbox{$\sqrt{\rm {\sl s}}$}}
\newcommand{\fdel}{\mbox{$f(\DEP)$}}
\newcommand{\fdele}{\mbox{$f(\DEP)^{exp}$}}
\newcommand{\fgap}{\mbox{$f(\DEP\! \geq \!3)$}}
\newcommand{\fgape}{\mbox{$f(\DEP\! \geq \!3)^{exp}$}}
\newcommand{\fpyt}{\mbox{$f(\DEP\!>\!2)$}}
\newcommand{\delth}{\mbox{$\DEP\! \geq \!3$}}
\newcommand{\uplim}{\mbox{$1.1\!\times\!10^{-2}$}}
\def\simge
{\mathrel{\rlap{\raise 0.53ex \hbox{$>$}}{\lower 0.53ex \hbox{$\sim$}}}}
\def\simle
{\mathrel{\rlap{\raise 0.53ex \hbox{$<$}}{\lower 0.53ex \hbox{$\sim$}}}}
\newcommand{\pbarp}{\mbox{$p\bar{p}$}}
\def\ETmiss{\mbox{${\hbox{$E$\kern-0.5em\lower-.1ex\hbox{/}\kern+0.15em}}_{\rm T}$}}
\def\Et{\mbox{$E_{T}$}}
\newcommand{\modeta}{\mid \!\! \eta \!\! \mid}
\def\gevcc{GeV/c$^2$}                   
\def\gevc{GeV/c}                        
\def\gev{GeV}                           
\newcommand{\als}{\mbox{${\alpha_{{\rm s}}}$}}
\def\1960{$\sqrt{s}=1960$ GeV}
\def\etI{E_{T_1}}
\def\etII{E_{T_2}}
\def\itaI{\eta_1}
\def\itaII{\eta_2}
\def\deta{\Delta\eta}
\def\etab{\bar{\eta}}
\def\xq{($x_1$,$x_2$,$Q^2$)}
\def\xx{($x_1$,$x_2$)}
\def\rap{pseudorapidity}
\def\as{\alpha_s}
\def\ap{\alpha_{\rm BFKL}}
\def\apb{\alpha_{{\rm BFKL}_{bin}}}
\def\cm{c.m.}

\def\MET{{\mbox{$E\kern-0.57em\raise0.19ex\hbox{/}_{T}$}}}
\def\METbold{{\mbox{$\mathbf{E\kern-0.57em\raise0.19ex\hbox{/}_{T}}$}}}

\newcolumntype{p}[0]{D{/}{\pm}{3,3}}
\newcolumntype{s}[0]{D{/}{/}{-1}}

\hspace{5.2in} \mbox{FERMILAB-PUB-10-320-E} 

\title{Search for events with leptonic jets and missing transverse energy in $\mathbf{p\bar{p}}$ collisions at $\mathbf{\sqrt{s}=1.96}$ TeV}

%
\affiliation{Universidad de Buenos Aires, Buenos Aires, Argentina}
\affiliation{LAFEX, Centro Brasileiro de Pesquisas F{\'\i}sicas, Rio de Janeiro, Brazil}
\affiliation{Universidade do Estado do Rio de Janeiro, Rio de Janeiro, Brazil}
\affiliation{Universidade Federal do ABC, Santo Andr\'e, Brazil}
\affiliation{Instituto de F\'{\i}sica Te\'orica, Universidade Estadual Paulista, S\~ao Paulo, Brazil}
\affiliation{Simon Fraser University, Vancouver, British Columbia, and York University, Toronto, Ontario, Canada}
\affiliation{University of Science and Technology of China, Hefei, People's Republic of China}
\affiliation{Universidad de los Andes, Bogot\'{a}, Colombia}
\affiliation{Charles University, Faculty of Mathematics and Physics, Center for Particle Physics, Prague, Czech Republic}
\affiliation{Czech Technical University in Prague, Prague, Czech Republic}
\affiliation{Center for Particle Physics, Institute of Physics, Academy of Sciences of the Czech Republic, Prague, Czech Republic}
\affiliation{Universidad San Francisco de Quito, Quito, Ecuador}
\affiliation{LPC, Universit\'e Blaise Pascal, CNRS/IN2P3, Clermont, France}
\affiliation{LPSC, Universit\'e Joseph Fourier Grenoble 1, CNRS/IN2P3, Institut National Polytechnique de Grenoble, Grenoble, France}
\affiliation{CPPM, Aix-Marseille Universit\'e, CNRS/IN2P3, Marseille, France}
\affiliation{LAL, Universit\'e Paris-Sud, CNRS/IN2P3, Orsay, France}
\affiliation{LPNHE, Universit\'es Paris VI and VII, CNRS/IN2P3, Paris, France}
\affiliation{CEA, Irfu, SPP, Saclay, France}
\affiliation{IPHC, Universit\'e de Strasbourg, CNRS/IN2P3, Strasbourg, France}
\affiliation{IPNL, Universit\'e Lyon 1, CNRS/IN2P3, Villeurbanne, France and Universit\'e de Lyon, Lyon, France}
\affiliation{III. Physikalisches Institut A, RWTH Aachen University, Aachen, Germany}
\affiliation{Physikalisches Institut, Universit{\"a}t Freiburg, Freiburg, Germany}
\affiliation{II. Physikalisches Institut, Georg-August-Universit{\"a}t G\"ottingen, G\"ottingen, Germany}
\affiliation{Institut f{\"u}r Physik, Universit{\"a}t Mainz, Mainz, Germany}
\affiliation{Ludwig-Maximilians-Universit{\"a}t M{\"u}nchen, M{\"u}nchen, Germany}
\affiliation{Fachbereich Physik, Bergische  Universit{\"a}t Wuppertal, Wuppertal, Germany}
\affiliation{Panjab University, Chandigarh, India}
\affiliation{Delhi University, Delhi, India}
\affiliation{Tata Institute of Fundamental Research, Mumbai, India}
\affiliation{University College Dublin, Dublin, Ireland}
\affiliation{Korea Detector Laboratory, Korea University, Seoul, Korea}
\affiliation{CINVESTAV, Mexico City, Mexico}
\affiliation{FOM-Institute NIKHEF and University of Amsterdam/NIKHEF, Amsterdam, The Netherlands}
\affiliation{Radboud University Nijmegen/NIKHEF, Nijmegen, The Netherlands}
\affiliation{Joint Institute for Nuclear Research, Dubna, Russia}
\affiliation{Institute for Theoretical and Experimental Physics, Moscow, Russia}
\affiliation{Moscow State University, Moscow, Russia}
\affiliation{Institute for High Energy Physics, Protvino, Russia}
\affiliation{Petersburg Nuclear Physics Institute, St. Petersburg, Russia}
\affiliation{Stockholm University, Stockholm and Uppsala University, Uppsala, Sweden }
\affiliation{Lancaster University, Lancaster LA1 4YB, United Kingdom}
\affiliation{Imperial College London, London SW7 2AZ, United Kingdom}
\affiliation{The University of Manchester, Manchester M13 9PL, United Kingdom}
\affiliation{University of Arizona, Tucson, Arizona 85721, USA}
\affiliation{University of California Riverside, Riverside, California 92521, USA}
\affiliation{Florida State University, Tallahassee, Florida 32306, USA}
\affiliation{Fermi National Accelerator Laboratory, Batavia, Illinois 60510, USA}
\affiliation{University of Illinois at Chicago, Chicago, Illinois 60607, USA}
\affiliation{Northern Illinois University, DeKalb, Illinois 60115, USA}
\affiliation{Northwestern University, Evanston, Illinois 60208, USA}
\affiliation{Indiana University, Bloomington, Indiana 47405, USA}
\affiliation{Purdue University Calumet, Hammond, Indiana 46323, USA}
\affiliation{University of Notre Dame, Notre Dame, Indiana 46556, USA}
\affiliation{Iowa State University, Ames, Iowa 50011, USA}
\affiliation{University of Kansas, Lawrence, Kansas 66045, USA}
\affiliation{Kansas State University, Manhattan, Kansas 66506, USA}
\affiliation{Louisiana Tech University, Ruston, Louisiana 71272, USA}
\affiliation{University of Maryland, College Park, Maryland 20742, USA}
\affiliation{Boston University, Boston, Massachusetts 02215, USA}
\affiliation{Northeastern University, Boston, Massachusetts 02115, USA}
\affiliation{University of Michigan, Ann Arbor, Michigan 48109, USA}
\affiliation{Michigan State University, East Lansing, Michigan 48824, USA}
\affiliation{University of Mississippi, University, Mississippi 38677, USA}
\affiliation{University of Nebraska, Lincoln, Nebraska 68588, USA}
\affiliation{Rutgers University, Piscataway, New Jersey 08855, USA}
\affiliation{Princeton University, Princeton, New Jersey 08544, USA}
\affiliation{State University of New York, Buffalo, New York 14260, USA}
\affiliation{Columbia University, New York, New York 10027, USA}
\affiliation{University of Rochester, Rochester, New York 14627, USA}
\affiliation{State University of New York, Stony Brook, New York 11794, USA}
\affiliation{Brookhaven National Laboratory, Upton, New York 11973, USA}
\affiliation{Langston University, Langston, Oklahoma 73050, USA}
\affiliation{University of Oklahoma, Norman, Oklahoma 73019, USA}
\affiliation{Oklahoma State University, Stillwater, Oklahoma 74078, USA}
\affiliation{Brown University, Providence, Rhode Island 02912, USA}
\affiliation{University of Texas, Arlington, Texas 76019, USA}
\affiliation{Southern Methodist University, Dallas, Texas 75275, USA}
\affiliation{Rice University, Houston, Texas 77005, USA}
\affiliation{University of Virginia, Charlottesville, Virginia 22901, USA}
\affiliation{University of Washington, Seattle, Washington 98195, USA}
\author{V.M.~Abazov} \affiliation{Joint Institute for Nuclear Research, Dubna, Russia}
\author{B.~Abbott} \affiliation{University of Oklahoma, Norman, Oklahoma 73019, USA}
\author{M.~Abolins} \affiliation{Michigan State University, East Lansing, Michigan 48824, USA}
\author{B.S.~Acharya} \affiliation{Tata Institute of Fundamental Research, Mumbai, India}
\author{M.~Adams} \affiliation{University of Illinois at Chicago, Chicago, Illinois 60607, USA}
\author{T.~Adams} \affiliation{Florida State University, Tallahassee, Florida 32306, USA}
\author{G.D.~Alexeev} \affiliation{Joint Institute for Nuclear Research, Dubna, Russia}
\author{G.~Alkhazov} \affiliation{Petersburg Nuclear Physics Institute, St. Petersburg, Russia}
\author{A.~Alton$^{a}$} \affiliation{University of Michigan, Ann Arbor, Michigan 48109, USA}
\author{G.~Alverson} \affiliation{Northeastern University, Boston, Massachusetts 02115, USA}
% ...
%
%
\collaboration{The D0 Collaboration\footnote{with visitors from
$^{a}$Augustana College, Sioux Falls, SD, USA,
$^{b}$The University of Liverpool, Liverpool, UK,
$^{c}$SLAC, Menlo Park, CA, USA,
$^{d}$ICREA/IFAE, Barcelona, Spain,
$^{e}$Centro de Investigacion en Computacion - IPN, Mexico City, Mexico,
$^{f}$ECFM, Universidad Autonoma de Sinaloa, Culiac\'an, Mexico,
and
$^{g}$Universit{\"a}t Bern, Bern, Switzerland.%
}} \noaffiliation
\vskip 0.25cm

\date{August 19, 2010}

\begin{abstract}
We present the first search for pair production of isolated jets of
charged leptons in association with a large imbalance in transverse
energy in $p\bar{p}$ collisions using 5.8~\ifb\ of integrated
luminosity collected by the D0 detector at the Fermilab Tevatron
Collider. No excess is observed above Standard Model background, and
the result is used to set upper limits on the production cross
section of pairs of supersymmetric chargino and neutralino particles
as a function of ``dark-photon'' mass, where the dark photon is
produced in the decay of the lightest supersymmetric particle.
\end{abstract}


\maketitle

\begin{figure}\centering
\includegraphics[height=5cm]{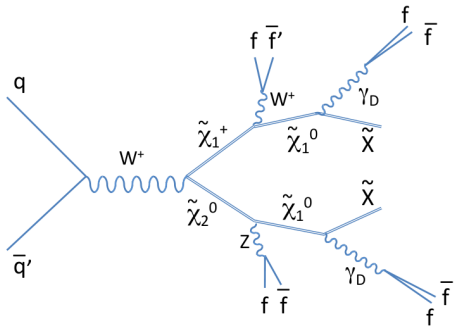}
\caption{A diagram for associated production of SUSY charginos and
neutralinos that decay into SM vector bosons and SM-LSPs
($\tilde{X}_1^0$), each decaying into the LSP of the hidden-sector
($\tilde{X}$) and a dark-photon ($\gamma_D$).} \label{fig:diagram1}
\end{figure}

Hidden-valley models~\cite{hv} contain a hidden sector that is very
weakly coupled to standard-model (SM) particles. By introducing new
low-mass particles in the hidden sector, these models have been
shown to provide cogent interpretation~\cite {idm, dmtheory} of
possible astrophysical anomalies~\cite{FermiLat, PAMELA, ATIC}, and
accommodate discrepancies in direct searches for dark matter
\cite{DAMA, CDMS}. The impact of the hidden valley particles should
be observable in high-energy collisions~\cite{itay, wacker, sundrum,
falkowski}. Although details of the hidden sector can affect the
phenomenology, the force carrier in the hidden sector, the
dark-photon ($\gamma_D$), must have a mass $\lesssim$ 2~GeV, and
generally decays into SM charged-fermion (or pion) pairs. In many
models, $\gamma_D$ has a short lifetime, and does not travel an
observable distance ($\lesssim$ 1 $\mu$m) before decaying. If
supersymmetry (SUSY) is realized in Nature, there will be partners
for both the SM and the hidden sector particles. If the lightest
SUSY particle (LSP) of the hidden sector ($\tilde{X}$) is lighter
than the lightest SM SUSY partner (SM-LSP), the SM-LSP can decay
promptly into particles of the hidden sector, and always will do so
if $R$-parity is conserved. The D0 collaboration has reported
\cite{d0dg} a search for such a decay, with one SM-LSP decaying to a
SM photon and $\tilde{X}$, and the other to $\gamma_D$ and
$\tilde{X}$. However, the SM-LSP might decay predominantly into
hidden sector particles, thereby yielding two or more $\gamma_D$ in
each event, as indicated in Fig.~\ref{fig:diagram1}. Pair-produced
dark photons could also arise from rare decays of $Z$
bosons~\cite{itay, Cheung:2009su} and Higgs bosons~\cite{falkowski}.
Single dark photons should also be produced directly in association
with a jet, as in SM prompt-photon production. This process is
difficult to detect at a hadron collider, while high-luminosity
low-energy $e^+e^-$ colliders could be more effective in observing
such events~\cite{dg-babar, dg-lowE}.

Since hidden-sector particles have small mass and they are produced
with high velocities, their decays through the hidden sector can
produce jets of tightly collimated particles from decays of
$\gamma_D$. If $M(\gamma_D)<2 m(\pi)$, the jets will consist only of
charged leptons. Even for larger $M(\gamma_D)$, the lepton content
of these jets will be high, and we therefore refer to them as
leptonic jets ($l$-jets). For the proposed scenario, every SUSY
event will have at least two $l$-jets and a large imbalance in
transverse energy (\MET) from the escaping $\tilde{X}$ and possibly
also from other escaping dark particles. Radiation of additional
$\gamma_D$ in the hidden sector \cite{itay} can dilute the $l$-jet
signatures, by producing final-state particles in $l$-jets that are
softer, less tightly collimated, and less isolated.

In this Letter, we present a search for events with two $l$-jets and
large \MET~in data collected using the D0~\cite{d0det} detector
during Run II of the Fermilab Tevatron Collider, corresponding to an
integrated luminosity of 5.8~\ifb. Depending on whether the
$\gamma_D$ decays to muons or electrons, the $l$-jet can appear
either as a ``muon $l$-jet'' or an ``electron $l$-jet'' in the
detector. To reconstruct muon $l$-jets, we demand a muon-track
candidate with hits in all three layers of the outer D0 muon system
and a matching track with $p_T>10$~GeV in the central tracker. An
electron $l$-jet must contain a central track with $p_T>10$~GeV that
matches an electromagnetic (EM) calorimeter cluster with transverse
energy $E_T^{\rm EM}>$15~GeV within a cone of radius ${\cal R} =
\sqrt{(\Delta\eta)^{2} + (\Delta\phi)^{2}} < 0.2$~\cite{d0coord}. EM
clusters are formed using a simple cone algorithm of ${\cal R} =
0.4$ and require $>95\%$ of the energy to be deposited in the EM
section of the calorimeter. The calorimeter isolation variable
${\cal I_{\it e}} = [E_T^{\text{tot}}(0.4) - E_T^{\rm
EM}(0.2)]/E_T^{\rm EM}(0.2)$ must be ${\cal I_{\it e}}<0.2$, where
$E_T^{\text{tot}}(0.4)$ is the total transverse energy in a cone of
radius ${\cal R}=0.4$, corrected for contributions from the
underlying event, and $E_T^{\rm EM}(0.2)$ is the transverse EM
energy in a cone of radius ${\cal R}=0.2$. The central ``seed''
track matched to the muon or EM cluster is required to have at least
one hit in the silicon detector. When the seed track is matched to
both a muon and an EM cluster, the $l$-jet is defined as a muon
$l$-jet. Next, a companion track of opposite electric charge from
the seed track, and within $z=1$~cm of the seed track at its
distance of closest approach to the beamline, is required to have
$p_T>4$~GeV and be within ${\cal R} < 0.2$ of the seed track. If
more than one such companion track is found, we use the one with
smallest ${\cal R}$. No explicit requirements are made on the
distances of closest approach of tracks to the collision point, thus
the $l$-jet reconstruction efficiency remains high for $\gamma_D$
decay radii up to $\approx$ 1 cm. We then choose the pair of $l$-jet
candidates with seed tracks separated by ${\cal R}>0.8$ that have
the largest invariant mass of any pair of seed tracks in the event.

The {\sc MadGraph}~\cite{mad} MC event generator, with P{\sc
YTHIA}~\cite{pythia} for showering and hadronization, is used to
simulate the signal, and these Monte Carlo (MC) events are then
processed through the full {\sc GEANT3}-based~\cite{geant}
D0-detector simulation and event reconstruction software. SUSY
events generated using SPS8~\cite{sps} parameters of the
gauge-mediated-SUSY-breaking (GMSB) model are used as a benchmark.
%
%
The efficiency to reconstruct many tightly-collimated tracks is
difficult to determine from data, and we therefore assume that all
neutralinos decay directly into a single $\gamma_D$ and the dark
gaugino LSP $\tilde{X}$, giving just two leptons per $l$-jet. The
$\tilde{X}$ would, most naturally, have a similar mass as
$\gamma_D$, so we assume $m(\tilde{X})=1$~GeV. More complicated
hidden-sector options are studied using MC simulation and are
discussed below.

The analysis requires two $l$-jet candidates (either muon or
electron) in each event. The three classes of $\mu\mu$, $e\mu$, and
$ee$ $l$-jets are analyzed separately, and contain 7344, 19014, and
30642 candidate events, respectively. Each event is assigned to just
one class, with preference of choice given to $\mu\mu$, then $e\mu$,
and then $ee$, since muon $l$-jets have less background. All
collected events are used in the analysis, but most pass single or
di-lepton triggers~\cite{d0det}. Following offline selections, the
trigger efficiency for signal is $>90\%$.

The main background to $l$-jets is from multijet production, but
electron $l$-jets also have a contribution from photon production
with subsequent conversion to $e^+e^-$. Such backgrounds cannot be
calculated reliably using simulation, and are therefore determined
from data. We exploit the tight collimation of $l$-jets to
distinguish them from multijet background, through track and
calorimeter-isolation criteria. The ``track isolation'' is defined
by a scalar sum over $p_T$ of tracks with $p_T>0.5$~GeV, $z<1$ cm
from the seed track at its distance of closest approach to the
beamline, and within an annulus $0.2<{\cal R}<0.4$ relative to the
seed track. Muon $l$-jet calorimeter isolation (${\cal I_{\mu}}$),
defined in Ref.~\cite{haa}, relies on the transverse energies of all
calorimeter cells within ${\cal R}<0.4$, excluding cells within
${\cal R}<0.1$ of either the seed muon or its companion track. For
electron $l$-jet isolation, we employ the EM cluster-isolation
${\cal I_{\it e}}$ defined above. A reliable estimate of background
requires that the $l$-jet isolation requirements not bias the
kinematics, such as distributions in \MET~or $p_T$ of $l$-jets. Both
types of $l$-jets require the track isolation to be ${\cal I_{\it
l}}<2$~GeV, which does not significantly bias the background.
Calorimeter-isolation criteria are chosen as linear functions of
$p_T$ values of the $l$-jet, such that the fraction of rejected
background is large, but weakly dependent on \MET, as discussed
below. For EM clusters, we choose ${\cal I}_{e} < 0.085 \times
p_T-0.53$ (in GeV units), which rejects 90\% of the background. For
muon $l$-jets we use the scalar sum of $p_T$ values of the muon and
companion tracks as a measure of $l$-jet $p_T$, and require ${\cal
I}_{\mu} < 0.066 \times p_T+2.35$ (in GeV units), which rejects 94\%
of the background. We compare the \MET~distribution in data with
just one isolated $l$-jet to those containing two (not necessarily
isolated) $l$-jets. The two distributions are observed to be very
similar, which indicates that the kinematic bias on \MET~from ${\cal
I_{\it e}}$ and ${\cal I_{\mu}}$ requirements is indeed small. We
therefore use the \MET~distribution in data without isolation
requirements as background for the data with two isolated $l$-jets,
since both samples are dominated by similar multijet processes.

\begin{figure}\centering
\includegraphics[height=5cm]{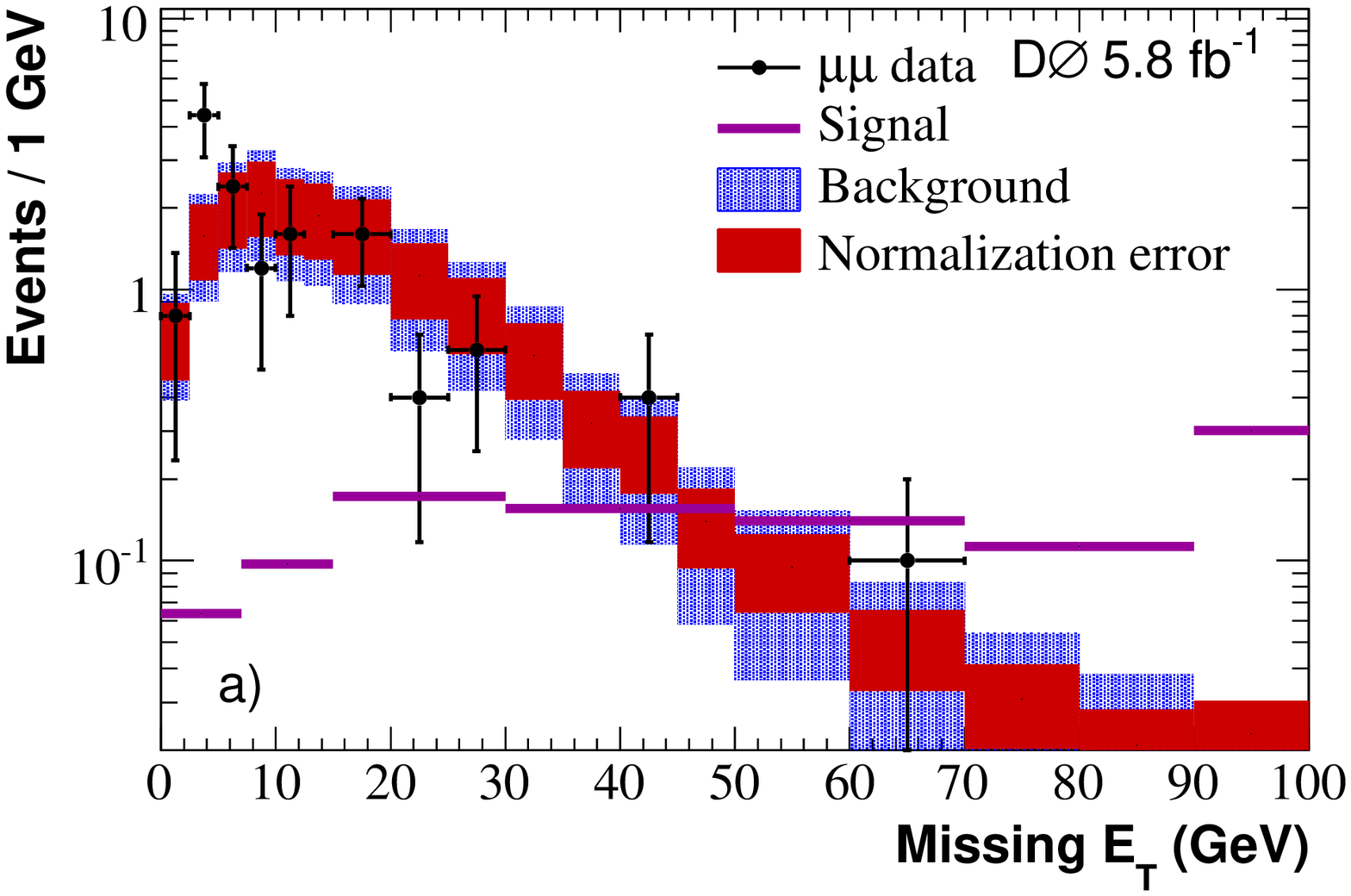}
\includegraphics[height=5cm]{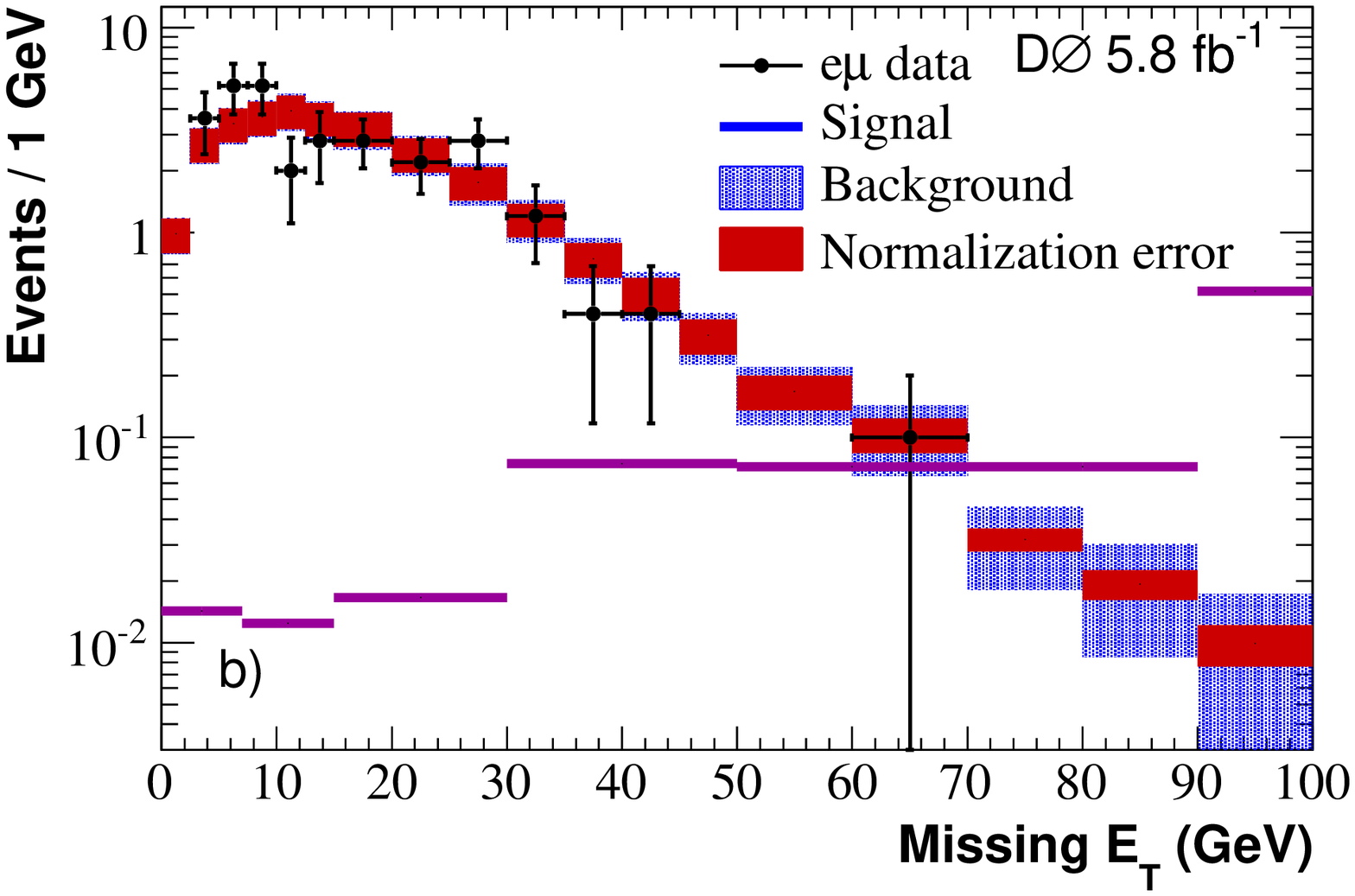}
\includegraphics[height=5cm]{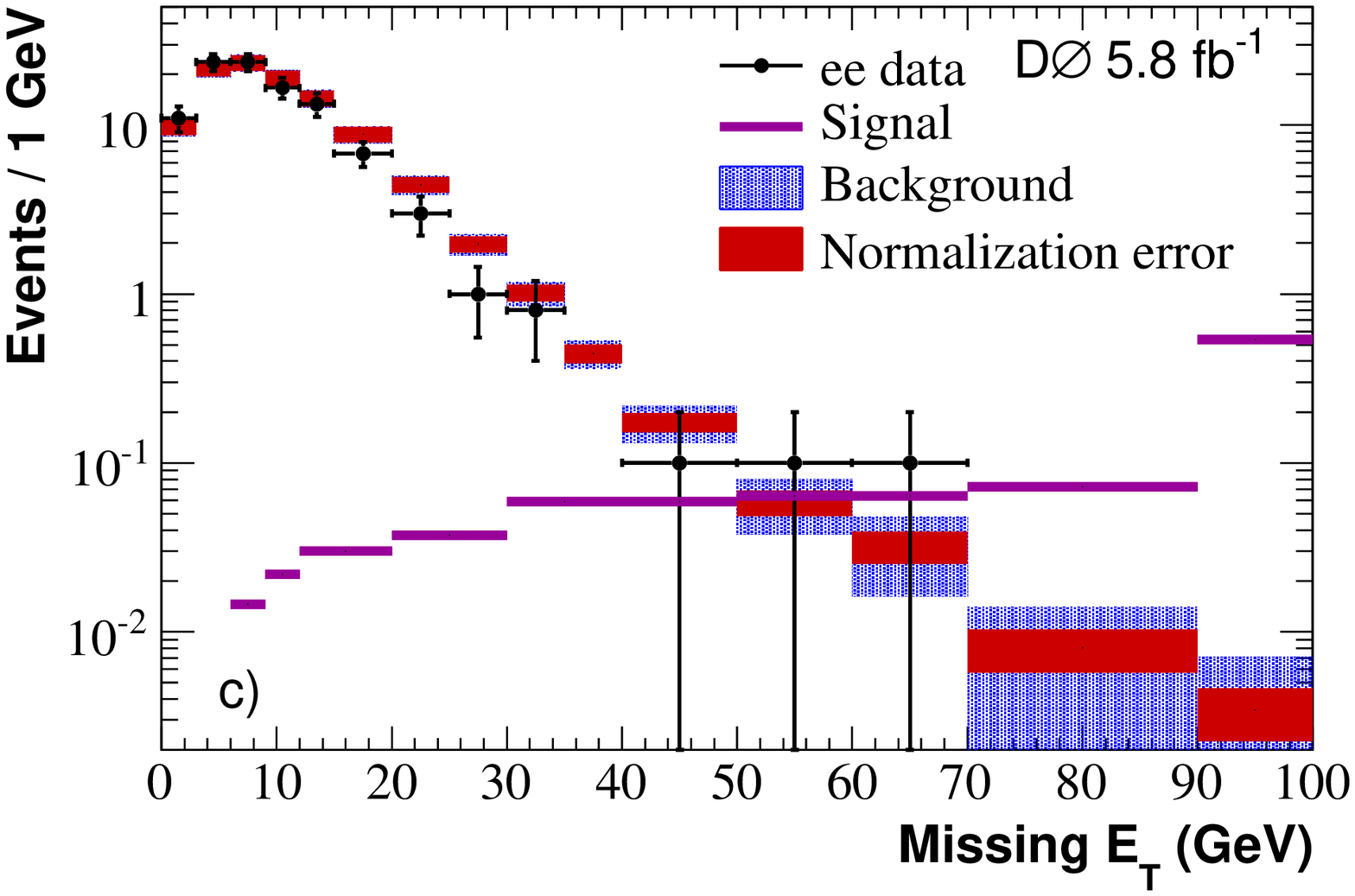}
\caption{(color online) The \MET~distribution for events with (a)
two isolated muon $l$-jets, (b) one muon and one electron $l$-jet,
and (c) two electron $l$-jets. The data are presented by the black
points, and the shaded bands represent the expected background, with
red showing the correlated part of the systematic uncertainty from
normalization and blue the full uncertainty. The SPS8 MC
contribution for signal (see text) is scaled to an integrated
content of 10 events. The highest bin contains all events with
\MET$>90$~GeV.} \label{fig:2lj}
\end{figure}

\begin{table} \centering
\caption{The ratio ${\cal R_s}$ of events with two $l$-jets and
$\MET>30$~GeV divided by the number with $\MET<15$~GeV in the
non-isolated data sample (see text); events observed and predicted
from background in each channel; the acceptance of the chosen
SPS8~\cite{sps} SUSY MC point, and the reconstruction efficiency,
given in \%; branching ratios (${\cal B}$) for each channel,
calculated from ${\cal B}_e$ and ${\cal B}_\mu$ in
Table~\ref{tab:masswindows}. Finally, limits on cross sections times
${\cal B}$ from the inclusive $l$-jet search.} \label{tab:acceff}
\begin{tabular}{cccpccc|c|c}
\hline\hline \multirow{2}{*}{Chan.} & \multirow{2}{*}{${\cal R_s}$}
& \multirow{2}{*}{$N_{\text{obs}}$ }&
\multicolumn{1}{c}{\multirow{2}{*}{$N_{\text{pred}}$}} &
\multirow{2}{*}{${\cal A}$(\%)} & \multirow{2}{*}{$\epsilon$(\%)}&
\multirow{2}{*}{${\cal B}$} &
\multicolumn{2}{c}{$\sigma_{95\%}\times {\cal B}$, fb }  \\
\cline{8-9}   &        &         &          &          &          &       & obs. & pred.  \\
\hline
$\mu\mu$ & 0.33 &   3 & 8.6/4.5 & 50 & 12 & ${\cal B}_{\mu}^2 $ & 20 & $35_{-21}^{+26}$ \\
$e\mu$   & 0.37 &  11 & 17.5/4.2 & 53 & 15 & $2{\cal B}_e {\cal B}_{\mu} $ & 19 & $30_{-15}^{+19}$ \\
$ee$     & 0.04 &  7 & 10.2/1.7 & 45 & 20 & ${\cal B}_e^2$ & 13 & $19_{-9}^{+11}$ \\
\hline\hline
\end{tabular}
\end{table}

Finally, we require $\MET>30$~GeV for the search sample, where
\MET~is calculated using only calorimetric information, and not
corrected for any detected muons, as muon reconstruction is
unreliable in $l$-jets because of the presence of nearby tracks. We
scale the \MET~distribution in the data sample without isolation
criteria so that the total number of events with $\MET<15$~GeV
matches that in the isolated data sample, see Fig.~\ref{fig:2lj}.
The ratio ${\cal R_s}$ defined as the number of events in each
search channel with $\MET>30$~GeV divided by the scaled number of
events with $\MET<15$~GeV in each respective background is given in
Table \ref{tab:acceff}. The value of ${\cal R_s}$ is important since
if a signal has a \MET~spectrum similar to that of the background,
this analysis would be largely insensitive, regardless of the size
of the signal. The total background for a signal having $f_1$ events
with \MET$<15$~GeV and $f_2$ events with \MET$>30$~GeV is a factor
of $(f_1/f_2)\times {\cal R_s}$ larger than for the case of no
signal. For the benchmark signals considered, $(f_1/f_2)\times {\cal
R_s} \ll 1$, and the correction is therefore ignored.

We separate the detection efficiency into three components (Table
\ref{tab:acceff}): (i) the branching ratio (${\cal B}$) for an event
to have at least two $l$-jets in the $\mu\mu$, $e\mu$, or $ee$
channel, obtained from the expected $\gamma_D$ branching fractions
\cite{d0dg}, (ii) the acceptance (${\cal A}$) for both $l$-jets to
have the seed and companion tracks within $|\eta|<1.1$ for electrons
and $<1.6$ for muons, with $p_T>$10 and 4~GeV, respectively, and
\MET~(calculated in MC as the vector sum of transverse momenta of
all stable particles in the hidden sector, neutrinos, and muons)
$>30$~GeV, and (iii) the efficiency ($\epsilon$) to reconstruct both
$l$-jets in the acceptance, to pass the isolation criteria for both
$l$-jets, and to have reconstructed \MET~in excess of 30~GeV. The
acceptance and reconstruction efficiency do not vary significantly
with $M(\gamma_D)$.

With no excess observed above the expected background at large
\MET~(see Fig.~\ref{fig:2lj}), we set limits on $l$-jet production
cross sections, using a likelihood fitter~\cite{collie} that
incorporates a log-likelihood ratio statistic~\cite{cls}. Limits at
the 95\% CL on cross section times ${\cal B}$, calculated separately
for the $\mu\mu$, $e\mu$, and $ee$ channels, using the observed
numbers of events, predicted backgrounds, and detection efficiencies
and acceptances, are given in Table \ref{tab:acceff}.
Systematic uncertainties are included for signal efficiency (20\%),
background normalization (20-50\%), and luminosity (6.1\%). The
uncertainty on the signal efficiency is dominated by the uncertainty
in the tracking efficiency for neighboring tracks in data. The
background uncertainty is dominated by the small remaining kinematic
bias on the \MET~arising from the isolation criteria.

\begin{table} \centering
\caption{Branching ratio (${\cal B}$) into electrons and muons of
$\gamma_D$ as a function of its mass. Mass windows for a search for
$\gamma_D$, and the efficiency for a reconstructed, isolated $l$-jet
to be found in each mass window, for electron and muon $l$-jets.}
\begin{tabular}{cccs}
\hline\hline
$M(\gamma_{D})$ (GeV) & ${\cal B}_e$/${\cal B}_\mu$ & $\Delta M$($l$-jet)(GeV) & \multicolumn{1}{c}{Eff.~$ee$/$\mu\mu$(\%)} \\
\hline
0.15      & 1.00/0.00 & 0.0--0.3 & 81/- \\
0.3       & 0.53/0.47 & 0.1--0.4 & 82/88 \\
0.5       & 0.40/0.40 & 0.3--0.6 & 81/89 \\
0.7       & 0.15/0.15 & 0.4--0.8 & 85/89 \\
0.9       & 0.27/0.27 & 0.6--1.1 & 82/91 \\
1.3       & 0.31/0.31 & 0.9--1.4 & 72/79 \\
1.7       & 0.22/0.22 & 1.0--1.8 & 73/76 \\
2.0       & 0.24/0.24 & 1.3--2.2 & 73/83 \\
\hline\hline
\end{tabular}
\label{tab:masswindows}
\end{table}

\begin{figure}\centering
\includegraphics[height=5cm]{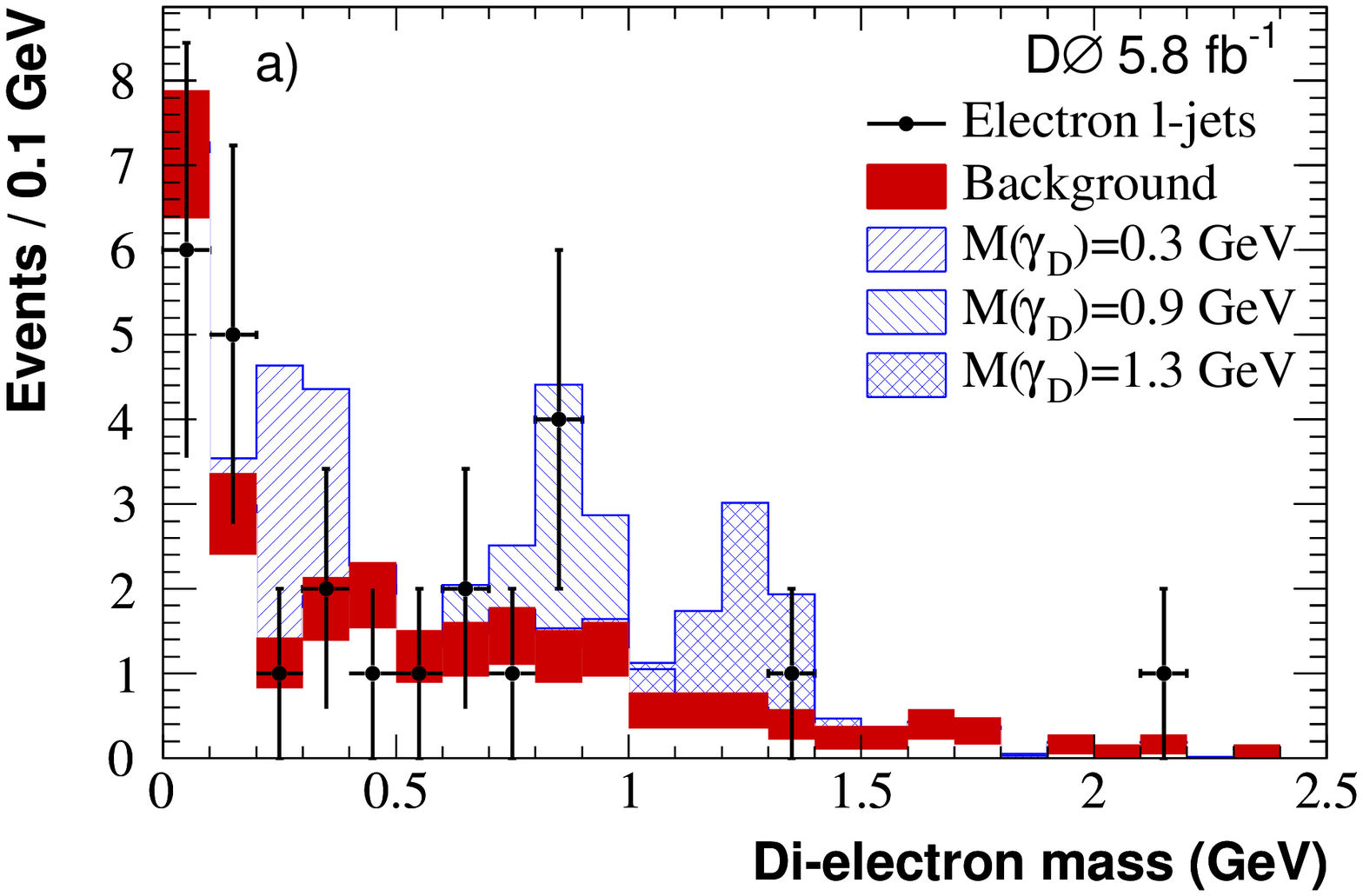}
\includegraphics[height=5cm]{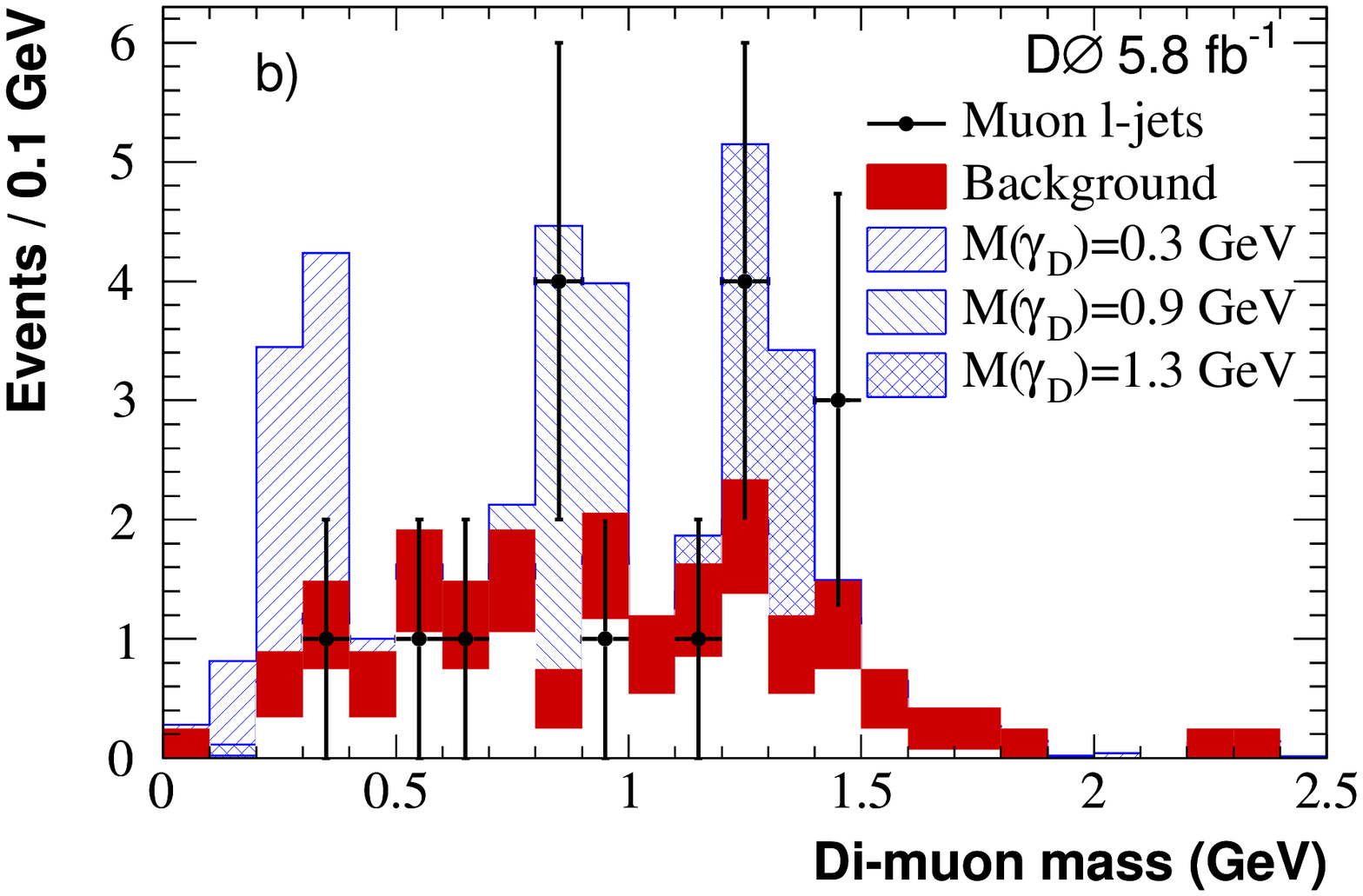}
\caption{(color online) Invariant mass of dark photon candidates
with two isolated $l$-jets and $\MET>30$~GeV, for (a) electron
$l$-jets (in the $ee$ and $e\mu$ channels) and (b) muon $l$-jets (in
the $e\mu$ and $\mu\mu$ channels). Each candidate event contributes
two entries, one for each $l$-jet. The red band shows the mass
distribution for events with $\MET<20$~GeV, normalized to the number
of entries with $\MET>30$~GeV. The shaded blue histograms show the
shapes of MC signals added to backgrounds, arbitrarily scaled to an
integrated content of 8 signal events, for $M(\gamma_D)$ = 0.3, 0.9,
and 1.3~GeV.} \label{fig:2_mass_lj}
\end{figure}

When the track multiplicity in any $l$-jet is small, the leading
track and its companion track are likely to originate from the decay
of the same dark photon, so we also examine the invariant mass of
the seed and its companion track ($M(\gamma_D)$) in events with two
isolated $l$-jets and \MET$>30$~GeV (Fig.~\ref{fig:2_mass_lj}). The
backgrounds are normalized by scaling the events passing all
selections but with $\MET<20$~GeV to data with $\MET>30$~GeV outside
of the mass windows defined in Tab.~\ref{tab:masswindows}, thus
${\cal R_s}$ is irrelevant for this second analysis. The selection
of background events is loosened to $\MET<20$~GeV for this resonance
search to increase the statistics of the sample. Limits on cross
sections are calculated in various ranges of $l$-jet mass, $\Delta
M$($l$-jet), as shown in Tab.~\ref{tab:masswindows} and
Fig.~\ref{fig:limit-inclusive}.

The dependence of the efficiency for reconstructing and identifying
$l$-jets on parameters of the hidden sector is studied using MC
simulation. Additional MC samples are used for examining the
neutralino decay into a dark Higgs boson that decays into two dark
photons, leading to more, but softer, leptons in $l$-jets.
Efficiency for these states
decreases by $\approx$50\% at large
$M(\gamma_D)$, for both electron and muon $l$-jets. The point
$M(\gamma_D)=0.7$~GeV also has a $\approx$50\% lower efficiency, due
to the large branching fraction of $\gamma_D$ to hadrons. MC events
are also generated with additional radiation in the hidden sector.
Raising the dark coupling ($\alpha_D$) from 0 to 0.3 reduces the
efficiency by up to 20\%, independent of $M(\gamma_D)$. According to
MC simulation, the $l$-jet identification criteria maintain good
efficiency even for more complicated behavior in the hidden sector.

In summary, we have performed a search for events with two tightly
collimated jets consisting mainly of charged leptons and large
\MET~in 5.8~\ifb\ of integrated luminosity. The invariant mass of
the $l$-jets, formed by a seed track and a companion track was also
examined for a resonant signal. No evidence was observed for such
signals, and upper limits were set, as a function of $M(\gamma_D)$,
on the production cross section for SUSY particles decaying to two
$l$-jets and large \MET.

\begin{figure}\centering
\includegraphics[height=5cm]{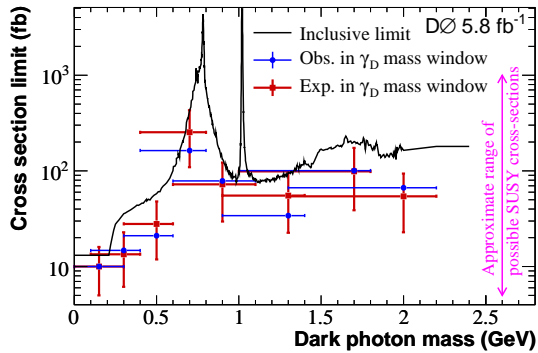}
\caption{(color online) Limit on the observed cross section (blue,
solid curve)
for the three channels combined, corrected for SPS8 acceptance, as a
function of $M(\gamma_D)$. Also shown are the observed (blue,
circles) and expected (red, squares) combined limit determined using
the measured masses of the seed and companion tracks in both
$l$-jets, for each mass window studied (from Table
\ref{tab:masswindows}). Limits are weaker when the dark photon
branching ratio to hadrons is larger, particularly near the $\rho$
and $\phi$ resonances.} \label{fig:limit-inclusive}
\end{figure}

We thank A. Falkowski, J. Ruderman, M. Strassler, S. Thomas, I.
Yavin, and J. Wacker for many useful discussions and guidance.
%
We thank the staffs at Fermilab and collaborating institutions,
and acknowledge support from the
DOE and NSF (USA);
CEA and CNRS/IN2P3 (France);
FASI, Rosatom and RFBR (Russia);
CNPq, FAPERJ, FAPESP and FUNDUNESP (Brazil);
DAE and DST (India);
Colciencias (Colombia);
CONACyT (Mexico);
KRF and KOSEF (Korea);
CONICET and UBACyT (Argentina);
FOM (The Netherlands);
STFC and the Royal Society (United Kingdom);
MSMT and GACR (Czech Republic);
CRC Program and NSERC (Canada);
BMBF and DFG (Germany);
SFI (Ireland);
The Swedish Research Council (Sweden);
and
CAS and CNSF (China).


\begin{thebibliography}{99}

\bibitem{hv}T. Han {\sl et al.},
J. High Energy Phys. {\bf 07}, 008 (2008); M. Strassler and K.
Zurek, Phys. Lett. B {\bf 651}, 374 (2007).

\bibitem{idm} D.P. Finkbeiner and N. Weiner, Phys. Rev. D {\bf 76} 083519  (2007).
\bibitem{dmtheory} N. Arkani-Hamed {\sl et al.},
Phys. Rev. D {\bf 79} 015014 (2009).

\bibitem{FermiLat} A.A. Abdo {\sl et al.}, Phys. Rev. Lett. {\bf 102}, 181101 (2009).
\bibitem{PAMELA} O. Adriani {\sl et al.}, Nature {\bf 458}, 607 (2009).
\bibitem{ATIC} J. Chang {\sl et al.}, Nature {\bf 456}, 362 (2008).
\bibitem{DAMA} R. Bernabei {\sl et al.} (DAMA/LIBRA Collaboration), Eur. Phys. J. C {\bf 56}, 333 (2008).
\bibitem{CDMS} Z. Ahmed {\sl et al.} (CDMS II Collaboration), Science {\bf 327} (5973), 1619 (2010).

\bibitem{itay} M. Baumgart {\sl et al.},
J. High Energy Phys. {\bf 04}, 014 (2009).
\bibitem{wacker} D.S.M. Alves {\sl et al.},
arXiv:0903.3945 [hep-ph] [Phys. Lett. B (to be published)].


\bibitem{sundrum} A. Katz and R. Sundrum,
J. High Energy Phys. {\bf 06}, 003 (2009).
\bibitem{falkowski} A. Falkowski {\sl et al.},
arXiv:1002.2952 [hep-ph].

\bibitem{d0dg} V. M. Abazov {\sl et al.} (D0 Collaboration), Phys. Rev. Lett. {\bf 103}, 081802, (2009).

\bibitem{Cheung:2009su}
  C.~Cheung {\sl et al.},
   J. High Energy Phys. {\bf 04}, 116 (2010).

\bibitem{dg-babar} B. Aubert {\sl et al.} (BaBar Collaboration),
arXiv:0908.2821; B. Aubert {\sl et al.} (BaBar Collaboration), Phys.
Rev. Lett. {\bf 103}, 081803 (2009).
\bibitem{dg-lowE} J. D. Bjorken {\sl et al.},
Phys. Rev. D {\bf 80} 075018 (2009).

\bibitem{d0det}
 V. M. Abazov {\it et al.}~(D0 Collaboration), Nucl.\ Instrum.\ Methods Phys.\
 Res.\ A {\bf 565}, 463 (2006).

\bibitem{d0coord}D0 uses a right-handed coordinate system, with the $z$-axis
pointing in the direction of the proton beam and the $y$-axis
pointing upwards. The azimuthal angle $\phi$ is defined in the $xy$
plane, and is measured from the $x$-axis. The pseudorapidity is
defined as $\eta = -\ln[\tan(\theta/2)]$, where $\theta$
is the polar angle.

\bibitem{mad}
 J.~Alwall {\it et al.},
   J. High Energy Phys. {\bf 09}, 028 (2007).

\bibitem{pythia}
 T.~Sj\"ostrand {\it et al.}, Comput.\ Phys.\ Commun.\ {\bf 135}, 238
 (2001).

\bibitem{geant}
 R. Brun and F. Carminati, CERN Program Library Long Writeup W5013,
1993 (unpublished).

\bibitem{sps}
The lightest neutralino mass for this SUSY point is $\approx$140~GeV
and the second neutralino and the chargino masses are both
$\approx$265~GeV;
B.C.~Allanach {\sl et al.}, Eur. Phys. J. C {\bf 25}, 113 (2002).

\bibitem{haa}
  V.~M.~Abazov {\it et al.}  (D0 Collaboration),
  Phys.\ Rev.\ Lett.\  {\bf 103}, 061801 (2009).


\bibitem{collie} W.~Fisher,
FERMILAB-TM-2386-E.

\bibitem{cls} T.~Junk, Nucl. Instrum. Methods Phys. Res. A {\bf 434}, 435 (1999);
A. Read, J. Phys. G {\bf 28}, 2693 (2002).


\end{thebibliography}
\end{document}